\documentclass[twocolumn,english]{revtex4}
\usepackage[T1]{fontenc}
\usepackage[latin9]{inputenc}
\usepackage{color}
\usepackage{array}
\usepackage{textcomp}
\usepackage{multirow}
\usepackage{amstext}
\usepackage{graphicx}

\makeatletter

\providecommand{\tabularnewline}{\\}

\@ifundefined{textcolor}{}
{%
 \definecolor{BLACK}{gray}{0}
 \definecolor{WHITE}{gray}{1}
 \definecolor{RED}{rgb}{1,0,0}
 \definecolor{GREEN}{rgb}{0,1,0}
 \definecolor{BLUE}{rgb}{0,0,1}
 \definecolor{CYAN}{cmyk}{1,0,0,0}
 \definecolor{MAGENTA}{cmyk}{0,1,0,0}
 \definecolor{YELLOW}{cmyk}{0,0,1,0}
}

\makeatother

\usepackage{babel}
\begin{document}
\title{\textcolor{black}{An efficient method for strongly correlated electrons
in one dimension}}
\author{Ion Mitxelena$^{1}$ and Mario Piris$^{1,2}$\bigskip{}
}
\address{$^{1}$Kimika Fakultatea, Euskal Herriko Unibertsitatea (UPV/EHU)
and Donostia International Physics Center (DIPC), P.K. 1072, 20080
Donostia, Euskadi, Spain. \\
$^{2}$Basque Foundation for Science (IKERBASQUE), 48013 Bilbao, Euskadi,
Spain.\bigskip{}
}
\begin{abstract}
The one-particle reduced density matrix functional theory in its natural
orbital functional (NOF) version is used to study strongly correlated
electrons. We show the ability of the Piris NOF 7 (PNOF7) to describe
non-dynamic correlation effects in one-dimensional (1D) systems. An
extensive study of 1D systems that includes Hydrogen (H) chains and
the 1D Hubbard model with periodic boundary conditions is provided.
Different filling situations and large sizes with up to 122 electrons
are considered. Compared to quasi-exact results, PNOF7 is accurate
in different correlation regimes for the 1D Hubbard model even away
from the half-filling, and maintains its accuracy when the system
size increases. The symmetric and asymmetric dissociations of the
linear H chain composed of 50 atoms are described to remark the importance
of long-range interactions in presence of strong correlation effects.
Our results compare remarkably well with those obtained at the density-matrix
renormalization group level of theory. 

\bigskip{}

\bigskip{}
\end{abstract}
\maketitle

\section{Introduction}

One-dimensional (1D) many-electron systems remain a non-trivial problem
for electronic structure methods. Density functional theory in its
conventional local or semilocal approximations is not able to provide
a correct description of correlated insulators \cite{Carrascal},
configuration interaction methods cannot deal with too large systems,
and coupled cluster singles and doubles with perturbative triples
(CCSD(T)) shows instabilities at large interatomic distances in 1D
chains of Hydrogen (H) atoms \cite{Hachman2006}. \textcolor{black}{Recently,
significant progress has been made in lattice density functional theory
\cite{Muller-lattice-dft-2019}. Nevertheless, the density-matrix
renormalization group (DMRG) algorithm \cite{white-prl-1992} remains
the most accurate method for studying 1D systems, including gapless
chains \cite{Papenbrock-dmrg-prb-2003,Cheranovskii_2019-jpcm-dmrg}.
Consequently, it will be employed as benchmark in this work.}

The electronic wavefunction is taken as a linear combination of geminal
functions to have a non-factorial scaling. In this context, variational
Monte Carlo calculations using a Jastrow-antisymmetrized geminal power
wavefunction has recently been used \cite{QMC_PRB} to successfully
investigate periodic 1D H chains. Another approach based on geminal
expansions is the antisymmetric product of 1-reference-orbital geminals
(AP1roG). The optimized orbital version of AP1roG (OO-AP1roG) has
proven \cite{Boguslawski2014} to be a reliable method for strongly
correlated 1D systems, such as the 1D Hubbard model with periodic
boundary conditions, as well as for metallic and molecular H rings.
Nevertheless, it has recently been shown \cite{Boguslawski2016} that
contributions from singly occupied states are important in the strong
correlation limit, so OO-AP1roG needs to include open-shell configurations
to accurately describe the $U/t\rightarrow\infty$ limit in the 1D
Hubbard model and the dissociation limit in H chains.

The natural orbital functional theory (NOFT) \cite{Piris2014a,Pernal2016}
constitutes an alternative to highly correlated methods. The energy
is expressed in terms of natural orbitals (NOs) and their occupation
numbers (ONs), so that from the outset NOFT correctly handles the
multiconfigurational character inherent in strongly correlated systems.
A route for the construction of an approximate natural orbital functional
(NOF) involves the employment of necessary $N$-representability conditions
for the two-particle reduced density matrix (2RDM) \cite{Piris2006}.
Appropriate 2RDM reconstructions have led to different implementations
known in the literature as PNOFi (i=1-7) \cite{piris2013ijqc,Piris2017}. 

The electron pairing approach came to the NOFT with the proposal of
PNOF5 \cite{Piris2011}. The latter is closely related to geminal
approaches, since it corresponds to an antisymmetrized product of
strongly orthogonal geminals \cite{Piris2013e}. PNOF5 draws a system
of $N$ electrons as independent electron pairs providing a good description
of the intrapair electron correlation, but lacks the correlation between
pairs. Consequently, a bad description of the strong correlation limit
is obtained \cite{mitxelena2018a}. To introduce interpair electron
correlation effects in singlet states, PNOF7 was proposed \cite{Piris2017,mitxelena2018a},
namely,
\begin{equation}
E={\displaystyle \sum_{g=1}^{N/2}}E_{g}+{\displaystyle \sum_{f\neq g}^{N/2}}E_{_{fg}}\label{PNOF7}
\end{equation}
where
\begin{equation}
\begin{array}{c}
E_{g}=2\sum\limits _{p\in\Omega_{g}}n_{p}\mathcal{H}_{pp}+\sum\limits _{q,p\in\Omega_{g}}\Pi_{qp}\mathcal{L}_{pq}\qquad\qquad\\
\Pi_{qp}=\left\{ \begin{array}{c}
\sqrt{n_{q}n_{p}}\,,\quad q=p\textrm{ or }q,p>\frac{N}{2}\\
-\sqrt{n_{q}n_{p}}\,,\quad q=g\textrm{ or }p=g\qquad\;
\end{array}\right.
\end{array}
\end{equation}
\begin{equation}
\begin{array}{c}
E_{fg}=\sum\limits _{p\in\Omega_{f}}\sum\limits _{q\in\Omega_{g}}\left[n_{q}n_{p}\left(2\mathcal{J}_{pq}-\mathcal{K}_{pq}\right)-\Phi_{q}\Phi_{p}\mathcal{L}_{pq}\right]\\
\Phi_{p}=\sqrt{n_{p}(1-n_{p})}.
\end{array}\label{Efg-Fi}
\end{equation}

$n_{p}$ stands for the ON of the spatial NO $\left|p\right\rangle $.
$\mathcal{H}_{pp}$ denotes the diagonal elements of the one-particle
part of the Hamiltonian involving the kinetic energy and the external
potential operators. $\mathcal{J}_{pq}$ and $\mathcal{K}_{pq}$ refer
to the usual Coulomb and exchange integrals, $\left\langle pq|pq\right\rangle $
and $\left\langle pq|qp\right\rangle $ respectively, whereas $\mathcal{L}_{pq}$
denotes the exchange-time-inversion integral $\left\langle pp|qq\right\rangle $. 

The orbital space is divided into\textit{ $N/2$} mutually disjoint
subspaces $\Omega{}_{g}$, so that $\sum_{p\in\Omega_{g}}n_{p}=1$.
Taking into account the spin, each $\Omega{}_{g}$ contains solely
an electron pair, and the normalization condition for the one-particle
reduced density matrix (1RDM) is automatically fulfilled: $2\sum_{p}n_{p}=N$.
Restriction of the ONs to the range $0\leq n_{p}\leq1$ represents
a necessary and sufficient condition for ensemble $N$-representability
of the 1RDM \cite{Coleman1963}.

It should be noted that $E_{g}$ reduces to the NOF obtained from
a two-electron singlet wavefunction, so the first term of PNOF7 accurately
describes the sum of electron-pair energies. The second term correlates
the motion of the electrons in different pairs with parallel and opposite
spins. For the latter, the particle-hole symmetry is explicitly considered
through $\Phi_{p}$ in the $\mathcal{L}$-term. This resembles the
original formulation of Bardeen, Cooper and Schrieffer (BCS) \cite{Bardeen1957},
which uses these types of interactions for all orbitals. The BCS method
is one of the best mean-field approaches to the Hubbard model with
attractive interactions \cite{Mingpu}, but underestimates the correlation
effects in systems with repulsive Hamiltonians \cite{Tsuchimochi_bcs}.
For the latter, recent studies \cite{Mitxelena2017a,mitxelena2018a,Mitxelena2018b}
suggest that PNOF7 could correctly recover the strong correlation
limit. In this paper, we provide an extensive study of H chains composed
of 50 atoms and the 1D Hubbard model in many filling situations, sizes,
and correlation regimes.

\textcolor{black}{The solution is established by optimizing the energy
(}\ref{PNOF7}\textcolor{black}{-\ref{Efg-Fi}) with respect to the
ONs and to the NOs, separately. The conjugate gradient method is used
to perform the optimization of the energy with respect to auxiliary
variables that enforce automatically the $N$-representability bounds
of the 1RDM. The self-consistent procedure proposed in \cite{Piris2009a}
yields the NOs by an iterative diagonalization procedure, in which
}orbitals are not constrained to remain fixed along the orbital optimization
process. All calculations have been carried out using the DoNOF code
developed in our group.

First, we show the ability of PNOF7 to describe the 1D Hubbard model.
The latter has the advantage of being extremely simple and is a useful
tool for benchmarking \cite{PHYSREVX_2Dhubbard}. The 1D Hubbard Hamiltonian
reads as
\begin{equation}
H=-t\sum_{\langle r,r'\rangle,\sigma}(a_{r,\sigma}^{\dagger}a_{r',\sigma}+a_{r',\sigma}^{\dagger}a_{r,\sigma})+U\sum_{r}n_{r,\alpha}n_{r,\beta}\label{HamiltonianHub}
\end{equation}
where $\langle r,r'\rangle$ indicates only near-neighbors hopping
between the sites $r$ and $r'$. $t>0$ is the hopping parameter
analogous to the kinetic energy, and $U$ is the electron-electron
on-site interaction parameter. $\sigma=\alpha,\beta$ stands for the
spin. $a_{r,\sigma}^{\dagger}\left(a_{r,\sigma}\right)$ is the creation
(annihilation) operator, so $n_{r,\sigma}=a_{r,\sigma}^{\dagger}a_{r,\sigma}$
gives the number of electrons on site $r$ with spin $\sigma$.

Let us restrict to the repulsive Hubbard model, hence $U$ is always
positive. $U/t$ is used as a dimensionless measure for the relative
contribution of both terms, therefore, at $U/t\rightarrow0$ (metallic
state) the mean-field theories work well due to the lack of two-electron
interactions, whereas at $U/t\rightarrow+\infty$ (insulating state)
strong correlations play the dominant role keeping electrons away
from each other.

\begin{figure}[t]
\centering{}\includegraphics[scale=0.9]{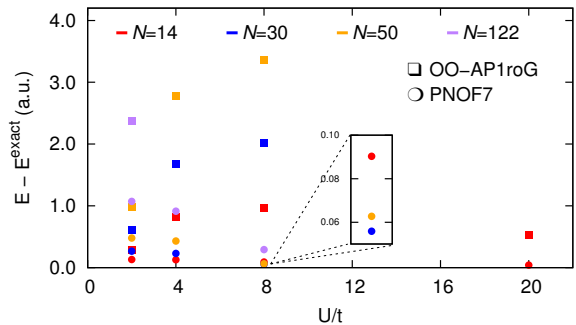}\caption{\label{fig4} Energy differences (a.u.) with respect to the exact
results for the 1D Hubbard model at half-filling with periodic boundary
conditions. OO-AP1roG and exact data from \cite{Boguslawski2014,Boguslawski2016}.
For $\mathrm{U}/\mathrm{t}=20$, only the result is reported for $N=14$.}
\end{figure}

In Fig. \ref{fig4}, we report the PNOF7 energy differences with respect
to the exact results for the 1D Hubbard model at half-filling. The
number of sites varies from 14 to 122 in small and intermediate correlation
regimes. For comparison, OO-AP1roG results \cite{Boguslawski2016}
have been included. The data sets used in this figure can be found
in the Supplemental Material (Table III). Note that OO-AP1roG deteriorates
for large systems (some errors fall out of Fig. \ref{fig4}), as well
as for large $U/t$ values. Conversely, PNOF7 is able to hold its
accuracy with respect to exact results when the system size increases.
For a given system, PNOF7 converges to the exact results in the strong
correlation limit.

Since the particle-hole symmetry is explicitly introduced into the
functional (\ref{PNOF7}-\ref{Efg-Fi}), PNOF7 is expected to be appropriate
for the half-filling case. Now we test the performance of PNOF7 away
from half-filling where the particle-hole symmetry is broken, so that
inhomogeneous phases can appear \cite{PHYSREVX_2Dhubbard}. The energy
per site for the 1D Hubbard model is shown in Table \ref{tab3}. We
focus on the strong correlation limit, i.e., large $U/t$ values,
which is particularly problematic for geminal-based theories like
OO-AP1roG \cite{Boguslawski2016}. For reference, we use the variational
2RDM (v2RDM) with P, Q and G $N$-representability constraints values
and quasi-exact results of the variational Matrix Product State (vMPS)
algorithm taken from Ref. \cite{VERSTICHEL201312}.

\begin{table}[t]
\begin{centering}
\caption{\label{tab3} Energy per site (a.u.) for 1D Hubbard model away from
half-filling at $U/t\rightarrow100$. Reference vMPS, v2RDM, and exact
data from \cite{VERSTICHEL201312}. $N_{sites}$ and $N$ stands for
the number of sites and electrons, respectively.\smallskip{}
}
\begin{tabular}{ccccccccccc}
$N_{sites}$ &  & $N$ &  & PNOF7 &  & vMPS &  & v2RDM &  & Exact{*}\tabularnewline
\hline 
\multirow{2}{*}{20} &  & \multirow{1}{*}{12} &  & -0.6025 &  & -1.0312 &  & -1.2177 &  & -1.0008\tabularnewline
 &  & \multirow{1}{*}{16} &  & -0.3820 &  & -0.4951 &  & -0.7860 &  & -0.4639\tabularnewline
\multirow{2}{*}{50} &  & \multirow{1}{*}{20} &  & -0.9081 &  & - &  & -1.2191 &  & -1.0008\tabularnewline
 &  & \multirow{1}{*}{40} &  & -0.4444 &  & - &  & -0.7862 &  & -0.4671\tabularnewline
\hline 
\end{tabular}\bigskip{}
\par\end{centering}
\raggedright{}{*}Exact results correspond to $U/t\rightarrow\infty$.
\end{table}

Table \ref{tab3} shows that PNOF7 remains close to vMPS for $N=16$
in 20 sites chain, whereas it lacks correlation energy for $N=12$.
In the case of 50 sites, PNOF7 produces accurate energies and it approaches
the exact result. Consequently, PNOF7 turns out to be particularly
accurate from a certain amount of electrons, from which the strong
correlation limit is described successfully. It is worth noting that
PNOF7 is more accurate than v2RDM when only two-particle conditions
are applied. It has recently been emphasized \cite{Mazziotti_Rubin2014,Verstichel_prl}
that three-particle conditions are needed in v2RDM to accurately describe
the strong correlation limit of the Hubbard model.

In a minimal basis set, there is only one band in 1D systems, therefore,
as long as long-range interactions are negligible, a linear chain
composed of H atoms resembles the 1D Hubbard model. Such a chain composed
of 50 H atoms is a simple prototype of strong correlation, and a challenging
test \cite{Hachman2006} for non-dynamic correlation.

\begin{figure}[b]
\centering{}\includegraphics[scale=0.6]{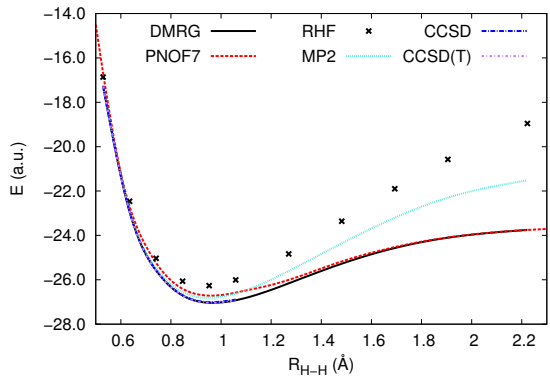}\caption{\label{fig1} Symmetric dissociation of linear $H_{50}$ using the
STO-6G basis set. RHF, MP2, CCSD, CCSD(T), and DMRG data from \cite{Hachman2006}.}
\end{figure}
\begin{table}[t]
\centering{}\caption{\label{tab1} Equilibrium distances ($R_{e}$) and dissociation energies
($D_{e}$) for the symmetric dissociation of linear $H_{50}$ using
the STO-6G basis set. RHF, MP2, PBE, OO-AP1roG, and DMRG data from
\cite{Boguslawski2014}.\bigskip{}
}
\begin{tabular}{c|c|c|c|c|c|c}
 & RHF & MP2 & PBE & OO-AP1roG & PNOF7 & DMRG\tabularnewline
\hline 
$R_{e}\,\left(\textrm{Å}\right)$ & 0.940 & 0.955 & 0.971 & 0.966 & 0.976 & 0.970\tabularnewline
$D_{e}\,\left(eV\right)$ & 199.0 & 144.1 & 146.6 & 82.2 & 86.9 & 89.7\tabularnewline
\end{tabular}
\end{table}

\begin{figure}[b]
\centering{}\includegraphics[scale=0.6]{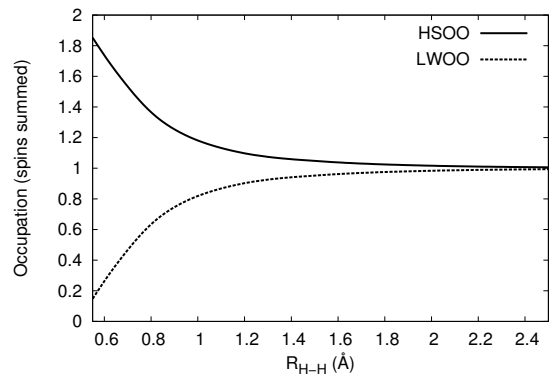}\caption{\label{fig2} ONs of the highest strongly occupied NO (HSOO) and the
lowest weakly occupied NO (LWOO) for the symmetric dissociation of
linear $H_{50}$ at the PNOF7/STO-6G level of theory.}
\end{figure}

In order to study the effect of long-range interactions, let us show
the ability of PNOF7 to describe bond-breaking processes. Fig. \ref{fig1}
shows the energies obtained for symmetric stretching of linear $H_{50}$
by using PNOF7, together with reference DMRG results and other well-established
electronic structure methods, namely, restricted Hartree-Fock (RHF),
second-order Möller-Plesset pertubation theory (MP2), CCSD, and CCSD(T).
All calculations were carried out using the STO-6G minimal basis \cite{emls}.
There is an outstanding agreement between PNOF7 and DMRG along the
dissociation curve, specially at large bond distances (insulating
phase) as well as at short H-H distances (metallic phase). At the
equilibrium distance, PNOF7 underestimates slightly the correlation,
however an inspection of spectroscopic constants (see Table \ref{tab1})
shows that PNOF7 agrees with DMRG better than standard methods such
as RHF, MP2, or the Perdew-Burke-Ernzerhof (PBE) density functional.
These methods fail dramatically at the dissociation limit \cite{Boguslawski2014}
since the occupancies become strongly fractional at intermediate and
long H-H distances, a behavior that PNOF7 (see Fig. \ref{fig2}) and
OO-AP1roG (see Fig. 4 in Ref. \cite{Boguslawski2014}) correctly reproduce.
Non-integer occupations also make CCSD and CCSD(T) not convergent
\cite{Hachman2006}, so the latter can be exclusively employed in
the equilibrium region. Note that OO-AP1roG underestimates the equilibrium
distance ($R_{e}$) and dissociation energy ($D_{e}$), whereas PNOF7
underestimates $D_{e}$ and yields slightly large $R_{e}$.

\begin{figure}
\centering{}\includegraphics[scale=0.6]{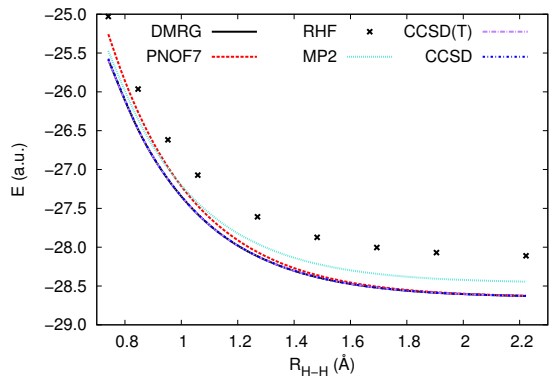}\caption{\label{fig3} Asymmetric dissociation of linear $H_{50}$ using the
STO-6G basis set. RHF, MP2, CCSD, CCSD(T), and DMRG data from \cite{Hachman2006}.}
\end{figure}

Fig. \ref{fig3} shows the energies obtained for the asymmetric dissociation
of linear $H_{50}$. It should be noted that the energy decreases
monotonically from the reference state composed of equidistant H atoms
to the set of independent $H_{2}$ molecules. In the asymmetric stretching,
we alternate the bond-stretching, so that half of the bonds remain
fixed, while the other half is stretched. In the dissociation limit,
we have 25 near-independent $H_{2}$ molecules. Similar to symmetric
dissociation, PNOF7 agrees with DMRG over large bond distances, whereas
there are slight differences at shorter bonds.

The results obtained for $H_{50}$ chains prove that numerical accuracy
of PNOF7 is comparable to that of the DMRG in many different correlation
regimes. This study includes the PNOF7 in the list of highly correlated
methods to study any system related to linear H chains \cite{review_Hchain}.

With the present work, a step forward has been taken in the development
of efficient methods for strong correlation. With a mean-field scaling,
the PNOF7 approximation compares with state-of-the-art methods for
describing strongly correlated electrons, e.g. DRMG, quantum Monte
Carlo or complete active space configuration interaction methods,
and overcomes the problems shown by similar approaches in the strong
correlation limit. The present paper will have a significant impact
on the development of new materials in which large unit cells are
required.

\textbf{\textcolor{black}{Acknowledgments:}}\textcolor{black}{{} Financial
support comes from MCIU/AEI/FEDER, UE (PGC2018-097529-B-100) and }Eusko
Jaurlaritza (Ref. IT1254-19)\textcolor{black}{. The authors thank
for technical and human support provided by IZO-SGI SGIker of UPV/EHU
and European funding (ERDF and ESF). }I.M. is grateful to Vice-Rectory
for research of the UPV/EHU for the Ph. D. grant (PIF//15/043).

\begin{table*}
\centering{}\caption{Energies (a.u.) for the 1D Hubbard model at half-filling with periodic
boundary conditions. OO-AP1roG, RHF, and exact data from Ref. \cite{Boguslawski2014,Boguslawski2016}.\smallskip{}
}
\begin{tabular}{ccccccccccc}
$N_{sites}$ &  & U/t &  & RHF &  & OO-AP1roG &  & PNOF7 &  & EXACT\tabularnewline
\hline 
\multirow{4}{*}{14} &  & 2 &  & -10.9758 &  & -11.6627 &  & -11.8230 &  & -11.9543\tabularnewline
 &  & 4 &  & -3.9758 &  & -7.2701 &  & -7.9610 &  & -8.0883\tabularnewline
 &  & 8 &  & 10.0242 &  & -3.6471 &  & -4.5228 &  & -4.6131\tabularnewline
 &  & 20 &  & 52.0242 &  & -1.4132 &  & -1.8932 &  & -1.9340\tabularnewline
\hline 
\multirow{3}{*}{30} &  & 2 &  & -23.2671 &  & -24.7779 &  & -25.1161 &  & \textminus 25.3835\tabularnewline
 &  & 4 &  & -8.2671 &  & -15.5495 &  & -17.0035 &  & \textminus 17.2335\tabularnewline
 &  & 8 &  & 21.7329 &  & -7.8152 &  & -9.78283 &  & \textminus 9.8387\tabularnewline
\hline 
\multirow{3}{*}{50} &  & 2 &  & -38.7039 &  & -41.2570 &  & -41.7650 &  & \textminus 42.2443\tabularnewline
 &  & 4 &  & -13.7039 &  & -25.9154 &  & -28.2696 &  & \textminus 28.6993\tabularnewline
 &  & 8 &  & 36.2961 &  & -13.0253 &  & -16.3215 &  & \textminus 16.3842\tabularnewline
\hline 
\multirow{3}{*}{122} &  & 2 &  & -94.3524 &  & \textminus 100.6497 &  & \textcolor{black}{-101.9499} &  & \textminus 103.0211\tabularnewline
 &  & 4 &  & -33.3524 &  & \textminus 63.2336 &  & -69.0861 &  & \textminus 70.0003\tabularnewline
 &  & 8 &  & 88.6476 &  & \textminus 31.7817 &  & -39.6698 &  & \textminus 39.9619\tabularnewline
\hline 
\end{tabular}
\end{table*}

\end{document}